\documentclass[%
reprint,
superscriptaddress,
amsmath,amssymb,
aps,
prb,
floatfix,
]{revtex4-1}

\usepackage{graphicx}
\usepackage{dcolumn}
\usepackage{bm}
\usepackage{xcolor}
\usepackage{ulem}
\usepackage{amsmath}

\newcommand{\add}[1]{{#1}}
\newcommand{\rmv}[1]{}

\begin{document}
	
	\preprint{APS/123-QED}
	
	\title{Effect of chemical substitution on the skyrmion phase in Cu$_2$OSeO$_3$}
	\author{Paul M. Neves}
	\affiliation{National Institute of Standards and Technology Center for Neutron Research, Gaithersburg, Maryland 20878, USA}
	\affiliation{University of Maryland, College Park, College Park, Maryland 20742, USA}
	\author{Dustin A. Gilbert}
	\affiliation{National Institute of Standards and Technology Center for Neutron Research, Gaithersburg, Maryland 20878, USA}
	\affiliation{Department of Materials Science and Engineering, University of Tennessee, Knoxville, Tennessee 37996, USA}
	\author{Sheng Ran}
	\affiliation{National Institute of Standards and Technology Center for Neutron Research, Gaithersburg, Maryland 20878, USA}
	\affiliation{University of Maryland, College Park, College Park, Maryland 20742, USA}
	\author{I-Lin Liu}
	\affiliation{National Institute of Standards and Technology Center for Neutron Research, Gaithersburg, Maryland 20878, USA}
	\affiliation{University of Maryland, College Park, College Park, Maryland 20742, USA}
	\author{Shanta Saha}
	\affiliation{University of Maryland, College Park, College Park, Maryland 20742, USA}
	\author{John Collini}
	\affiliation{National Institute of Standards and Technology Center for Neutron Research, Gaithersburg, Maryland 20878, USA}
	\affiliation{University of Maryland, College Park, College Park, Maryland 20742, USA}
	\author{Markus Bleuel}
	\affiliation{National Institute of Standards and Technology Center for Neutron Research, Gaithersburg, Maryland 20878, USA}
	\author{Johnpierre Paglione}
	\affiliation{University of Maryland, College Park, College Park, Maryland 20742, USA}
	\author{Julie A. Borchers}
	\affiliation{National Institute of Standards and Technology Center for Neutron Research, Gaithersburg, Maryland 20878, USA}
	\author{Nicholas P. Butch}
	\affiliation{National Institute of Standards and Technology Center for Neutron Research, Gaithersburg, Maryland 20878, USA}
	\affiliation{University of Maryland, College Park, College Park, Maryland 20742, USA}
	
	\date{\today}

	\begin{abstract}
		Magnetic skyrmions have been the focus of intense research due to their unique qualities which result from their topological protections. Previous work on Cu$_2$OSeO$_3$, the only known insulating multiferroic skyrmion material, has shown that chemical substitution alters the skyrmion phase. We chemically substitute Zn, Ag, and S into \add{powdered} Cu$_2$OSeO$_3$ to study the effect on the magnetic phase diagram. In both the Ag and the S substitutions, we find that the skyrmion phase is stabilized over a larger temperature range, as determined via magnetometry and small-angle neutron scattering (SANS). Meanwhile,
		while previous magnetometry characterization suggests two \add{high temperature} skyrmion phases in the Zn-substituted sample, SANS reveals\rmv{ that only} the high temperature phase \add{to be}\rmv{ is} skyrmionic while \add{we are unable to distinguish} the other\rmv{ cannot be distinguished} from helical order. Overall, chemical substitution weakens \add{helical and} skyrmion order \add{as inferred from neutron scattering of the $|q| \approx 0.01\ \text{\AA}^{-1}$ magnetic peak}\rmv{ in favor of disordered helices}.
	\end{abstract}
	
	
	\maketitle

	\section{Introduction}
	The study of quantum materials focuses on understanding the emergent properties of interacting electronic systems, including effects such as topology, and leveraging those properties to develop new technologies. Magnetic skyrmions are one such topologically protected spin arrangement;\cite{muhlbauer2009skyrmion,yu2010real,rossler2006spontaneous} the topological nature of skyrmions makes it extremely challenging to change the configuration, and thus makes them promising for data storage technologies\cite{fert2013skyrmions} with a high resistance to data corruption. Additionally, the relatively strong Dzyaloshinskii-Moriya interactions and direct spin interactions make these materials interesting from a fundamental condensed matter magnetism perspective.\cite{rossler2006spontaneous}
	
	One especially interesting skyrmion material is Cu$_2$OSeO$_3$ \cite{seki2012observation,adams2012long,onose2012observation}. Being the only known insulating, skyrmion material makes it particularly attractive for low energy and high frequency spintronic devices \cite{white2012electric,jiang2015blowing}, but the temperature and field range of the skyrmion phase is small, occurring at T $\approx$ 57 K $\pm$ 1 K, and H $\approx$ 18 mT $\pm$ 6 mT. Therefore, it is desirable to expand the skyrmion stability envelope.
	
	Chemical substitution has been previously demonstrated as an approach to improve the skyrmion stability. Ni substitution on the Cu site shows an enhanced skyrmion temperature range \cite{chandrasekhar2016effects} while Te substitution of Se shows a reduction in the skyrmion temperature stability \cite{wu2015physical}. Distinct from other substitutions, Zn substitution on the Cu site reportedly splits the skyrmion phase into two discontinuous temperature ranges \cite{wu2015unexpected}. \add{In contrast to recent reports of a second, low temperature skyrmion phase in the parent compound \cite{chacon2018observation, bannenberg2019multiple, qian2018new},} later results demonstrated that \add{the signature of a second skyrmion phase in Zn substitutions}\rmv{ this} is caused by a coexistence of two distinct stoichiometric levels of Zn\rmv{ substitution}, where the nominally higher Zn concentration regions of the powder sample have a lower magnetic transition temperature, and thus a lower temperature candidate skyrmion region \cite{vstefanvcivc2018origin}. The splitting of the skyrmion phase regions reported in these works was inferred from DC susceptibility measurements, but no direct measurements of the magnetic configuration, such as Lorentz transmission electron microscopy (LTEM) or \add{small angle neutron scattering (}SANS\add{)}, confirmed the existence of two skyrmion phases \cite{wu2015unexpected, vstefanvcivc2018origin}. The challenge which has hindered SANS studies \add{of Zn substituted Cu$_2$OSeO$_3$} has been that these works were performed on powders, with each grain possessing its own skyrmion lattice, such that the resultant SANS pattern is a generic ring.
	
	Here, we present our study of new \add{powdered} substitutions of Ag and S, and previously studied \add{powdered} Zn substitutions using magnetization and \rmv{small angle neutron scattering (}SANS\rmv{)}. \add{From magnetization,} Zn, Ag, and S chemical substitutions all show enhanced skyrmion temperature stability with increased substitution. However, despite the expanded temperature range, SANS measurements suggest that the skyrmion phase becomes increasingly disordered with substitution. Additionally, as a split phase is striking, we directly probe for the presence of skyrmions within (Cu$_{1-x}$Zn$_x$)$_2$OSeO$_3$ using a SANS rotation technique reported previously \cite{gilbert2019precipitating}. These results \add{are consistent with two distinct stoichiometries within the sample, and identify a skyrmion phase in the higher temperature stoichiometry. However, we are unable to distinguish a second, lower temperature skyrmion phase.}\rmv{ show that while there is evidence for two separate stoichiometric phases within the sample, the ordered skyrmion phase is only present within the higher temperature stoichiometric phase.}

	\section{Experimental Methods}
	
	Powdered samples of chemically substituted Cu$_2$OSeO$_3$ were prepared following previously published techniques \cite{bos2008magnetoelectric}. The constituent powders [Alfa Aesar CuO, Puratronic, 99.995\% (metals basis) powder; Alfa Aesar SeO$_2$, Puratronic, 99.999\% (metals basis) Powder/Lump; Alfa Aesar AgO, 99.9\% (metals basis); Alfa Aesar CuS, 99.8\% (metals basis) -200 Mesh Powder; and ESPI Metals ZnO 99.999\% -200 Mesh Powder] were ground and mixed in the correct stoichiometric ratios before being sealed in evacuated quartz ampuoules. The samples were then sintered at 600 $^\circ$C for several days in a box furnace. Each sample was then crushed using a mortar and pestle, pressed into a pellet, and sintered two more times. Phase purity was confirmed in each sample with X-ray powder diffraction \add{(X-ray data shown in Appendix D)}.
	
	\begin{figure}
		\includegraphics[width=\columnwidth]{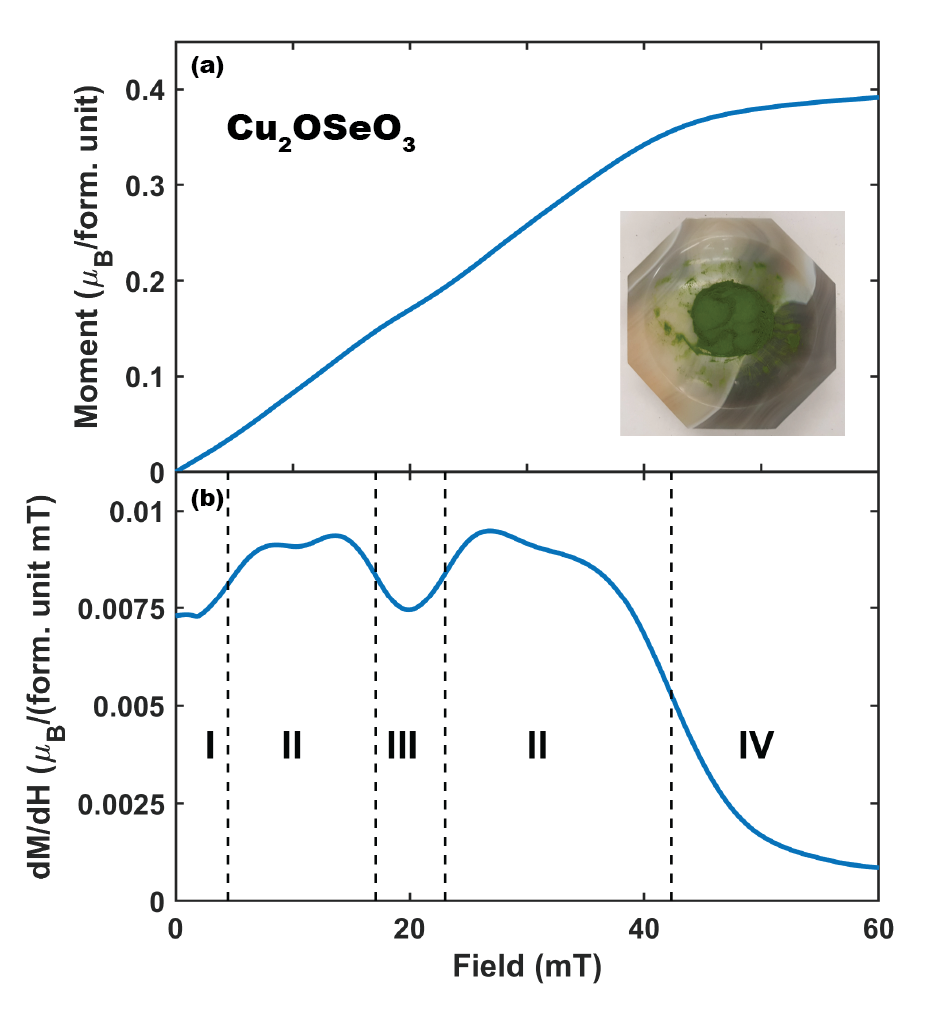}
		
		\caption{\label{fig:parent1DMPMS} Characteristic bulk magnetization for Cu$_2$OSeO$_3$ in the skyrmion phase. \textbf{(a)} Magnetization as a function of field for polycrystalline parent Cu$_2$OSeO$_3$ at 57 K. A photograph of the sample is included (inset). \textbf{(b)} The derivative $dM/dH$. Dashed lines mark approximate phase boundaries as determined by inflection points in $dM/dH$. Phase I corresponds to a multi-q helical structure, phase II represents a single-q helical structure, phase III is a Bloch-type skyrmion structure, and phase IV is a field-polarized state.}
	\end{figure}
	
	Magnetization measurements were conducted with a superconducting quantum interference device (SQUID) magnetometer at fixed temperatures while stepping the field between -4 mT and 120 mT. Field scans were performed in 1 K increments surrounding the skyrmion region of interest. The skyrmion phase can be identified by a double-inflection in the DC susceptibility data \cite{seki2012observation,adams2012long,onose2012observation,wu2015unexpected}, and measuring both temperature and field steps allows 2D contour mapping of the phase diagram.
	
	Neutron scattering measurements were performed at the NIST Center for Neutron Research on the NG-7 30 m SANS instrument. Approximately 50 mg of powdered sample was sealed in an aluminum foil packet and then placed in the beam. Absolute scattering intensity was calculated from empty beam flux. Temperatures were controlled with a closed cycle refrigerator (CCR), and magnetic fields were applied along the neutron flight path with an open bore resistive magnet. Scattering measurements were performed after decreasing the field from saturation \add{at $\approx$ 0.3 T}\rmv{ to remove hysteresis effects in the sample}. To resolve the skyrmion phase at selective fields and temperatures, the sample was rotated about the vertical axis from 0 degrees to $+$90 degrees to $-$90 degrees to 0 degrees approximately five times in the static magnetic field by manually turning the CCR in the magnet, and then measuring the SANS pattern.
	All measurement uncertainties, unless otherwise stated, reflect a standard uncertainty of one standard deviation.

	\section{Results and Discussion}
	
	\subsection{Magnetization and SANS}
\begin{figure}[h!]
		\includegraphics[width=\columnwidth]{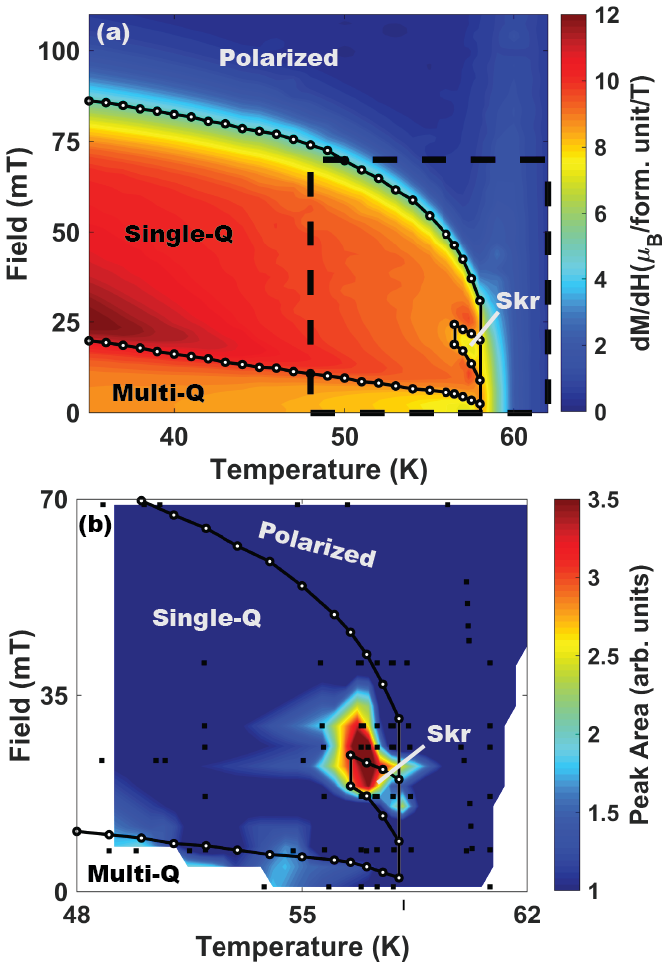}
		\caption{\label{fig:parentchar} Map of magnetic phase diagram from bulk magnetization and SANS. \textbf{(a)} $dM/dH$ [see, for example, Fig. \ref{fig:parent1DMPMS}(b)] for polycrystalline parent Cu$_2$OSeO$_3$ as a function of temperature and magnetic field. Phase boundaries are indicated with black lines [determined as shown in Fig. \ref{fig:parent1DMPMS}(b)]. \textbf{(b)} The integrated intensity of magnetic SANS scattering at the $|q| \approx 0.01\ \text{\AA}^{-1}$ peak associated with helimagnetic order. As the applied field lies parallel to the neutron beam, this map measures the helimagnetic order of the sample in the plane perpendicular to the applied field. The intense low-field and mid-field regions correspond to the multi-q helical and skyrmion phases, respectively, and agree with the phase boundaries established by the magnetization measurements from \textbf{(a)} (shown with black outlines). The locations of the SANS measurement points are indicated with black squares.}
	\end{figure}

	The magnetization of Cu$_2$OSeO$_3$ taken at 57 K is shown in Fig. \ref{fig:parent1DMPMS}(a), and the derivative ($dM/dH$, where $M$ is the magnetization and $H$ is the applied field) in Fig. \ref{fig:parent1DMPMS}(b); the skyrmion phase is readily identifiable in the derivative \cite{seki2012observation,adams2012long,onose2012observation,wu2015unexpected}. Starting from $H=0$, the magnetization steadily increases as the ``multi-q" helical state, labeled I in the derivative plot Fig. \ref{fig:parent1DMPMS}(b), deforms in response to the applied field. In this phase, helical propagation vectors are determined predominantly by magnetocrystalline anisotropy and shape anisotropy, generally preferring the \{100\} axes \cite{adams2012long,makino2017thermal}. However, at small fields, there is a transition from a ``multi-q" helical phase to a ``single-q" helical phase (labeled II), which is marked by a sharp increase in $dM/dH$ as the helices polarize along the field direction. The derivative then decreases again as the system enters the skyrmion phase (labeled III) as skyrmions form in the plane perpendicular to the applied field\rmv{, making further alignment more difficult}. The system then re-emerges from the skyrmion phase into the single-q helical phase with an increase in $dM/dH$. Finally, at high fields, the spins become field-polarized (labeled IV), and $dM/dH$ drops to near zero. These features allow rapid, rough identification of the magnetic phase boundaries. Indeed, plotting $dM/dH$ for all of the temperatures and fields in a contour plot, Fig. \ref{fig:parentchar}(a), clearly emphasizes each distinct phase.
	
	Spatially-resolved techniques are required to truly characterize the nature of the magnetic order in these compounds. Standard techniques are SANS \cite{muhlbauer2009skyrmion,white2012electric,seki2012formation,tokunaga2015new}, LTEM \cite{yu2010real,seki2012observation}, and X-ray microscopy and scattering \cite{woo2016observation,langner2014coupled,yamasaki2015dynamical,zhang2016resonant}. However, these techniques traditionally require either large, high quality single crystals as in SANS, or thin films or flakes as in LTEM. Realizing samples with controlled chemical substitution is challenging in single crystals \cite{vstefanvcivc2018origin}, but can be more easily achieved in polycrystalline samples \cite{chandrasekhar2016effects,vstefanvcivc2018origin,wu2015physical,wu2015unexpected}. This necessitates the development of alternative characterization techniques which are able to resolve\rmv{ helical order independent of the} \add{different types of helimagnetic order in the presence of} structural disorder.
	
	In a single crystal of Cu$_2$OSeO$_3$, skyrmions form in a 2D hexagonal lattice in the plane perpendicular to applied magnetic field with a skyrmion-skyrmion spacing of $\lambda \approx 70$ nm. This lattice generates a sixfold symmetric SANS pattern with $|q| \approx 0.01\ \text{\AA}^{-1}$ when measured with the incident neutrons parallel to the magnetic field \cite{chacon2018observation}. Surrounding the skyrmion phase in field and temperature lies the single-q helical phase. This phase consists of a helical vector with a pitch similar to the skyrmion phase, $\lambda \approx 70$ nm (as the order of the length scale is set by the same interaction energies \cite{rossler2011chiral, rossler2011chiral}), aligned along the direction of applied magnetic field. This creates a two-fold symmetric SANS pattern also with $|q| \approx 0.01\ \text{\AA}^{-1}$, but along the magnetic field axis \cite{chacon2018observation}. In the multi-q helical phase, helices orient along the family of \{001\} axes of the crystal, creating peaks on each crystal axis \cite{adams2012long}.
	
	In a polycrystalline skyrmion material, the SANS pattern is a sum of hexagons all with the same $|q|$ of $\approx 0.01\ \text{\AA}^{-1}$, but random orientations within the plane orthogonal to the magnetic field, resulting in a ring pattern. In contrast, the single-q helical phase remains relatively unchanged, as its symmetry is set by the magnetic field, not the crystal lattice. As such, there is a qualitative difference in SANS measurements between the skyrmion and single-q helical phase even in a polycrystalline sample. In contrast, the low-field, \textit{multi}-q helical phase shows a peak in scattering at $|q| \approx 0.01\ \text{\AA}^{-1}$ in the plane perpendicular to the magnetic field with no azimuthal dependence about the magnetic field. Thus, \textit{multi}-q helices and skyrmion lattices appear similar when the magnetic field is aligned with the neutron beam.
	
	\begin{figure}
		\includegraphics[width=\columnwidth]{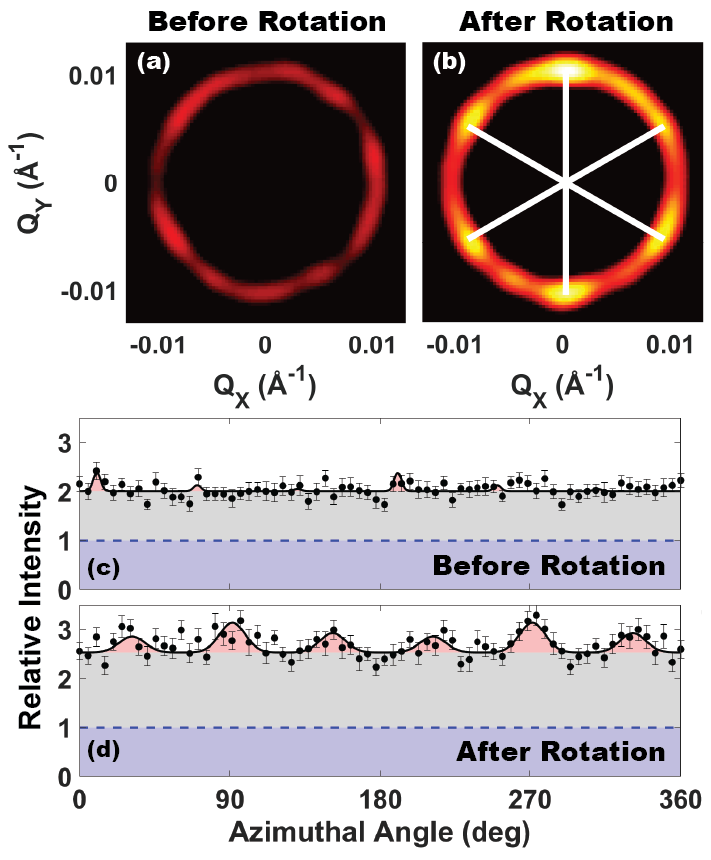}
		
		\caption{\label{fig:rotationGuide} Effect of rotation of skyrmions in a magnetic field on SANS pattern as reported in \cite{gilbert2019precipitating}. \textbf{(a)} SANS pattern of polycrystalline parent Cu$_2$OSeO$_3$ at approximately 57 K and 20 mT in the skyrmion phase (nonmagnetic structural scattering has been subtracted). No hexagonal symmetry is evident, as is shown in the corresponding annular plot averaging over $q = 0.0108 \pm 0.0024\ \text{\AA}^{-1}$ in \textbf{(c)}. \textbf{(b)} SANS signal after sample rotation (described in the text) in the 20 mT field. Hexagonal symmetry is clearly apparent (emphasized with white lines), and is further shown in the corresponding annular cut averaging over $q = 0.0108 \pm 0.0024\ \text{\AA}^{-1}$ in \textbf{(d)}.  Both \textbf{(c)} and \textbf{(d)} show the data (black points) and fits (black lines) to Equation \ref{eq:rotFit}, emphasizing how rotation enhances six-fold symmetric skyrmion lattice SANS scattering (red) on top of azimuthally uniform helimagnetic scattering (grey). The intensity in \textbf{(c)} and \textbf{(d)} is normalized to the nuclear scattering (blue). Details on the determination of each contribution to scattering given in Appendix B. \add{Both scattering patterns were integrated for 300 s immediately after rotation.}}
	\end{figure}
	
	\begin{figure*}
		\includegraphics[width=\textwidth]{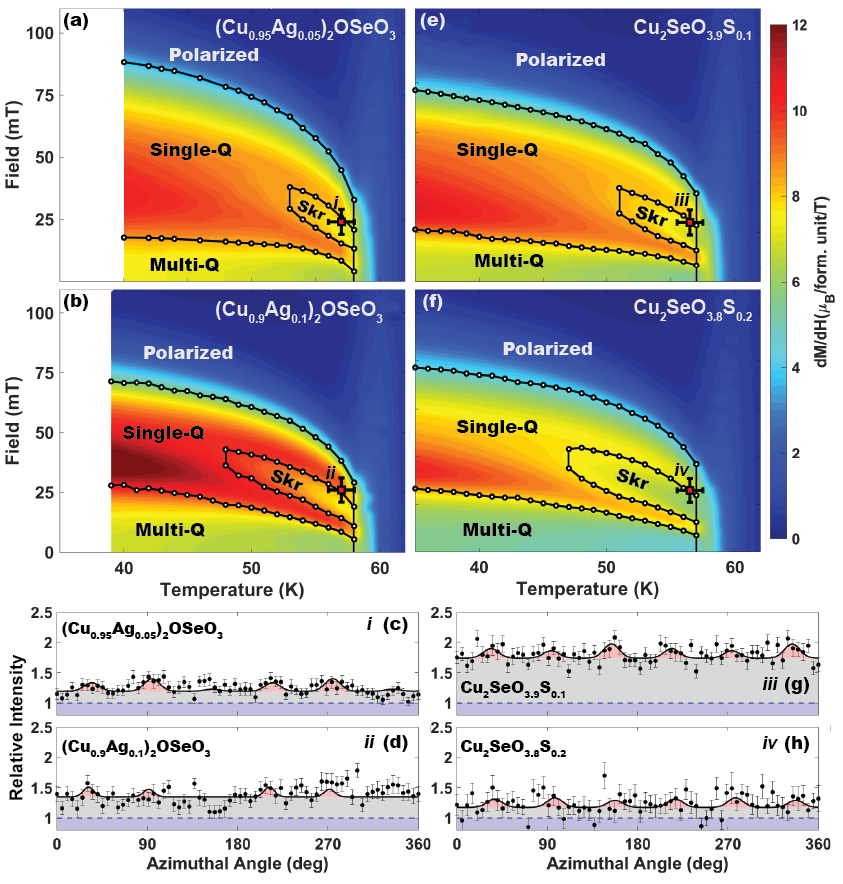}
		
		\caption{\label{fig:SAgMPMS} $dM/dH$ contours maps and SANS rotation measurements of polycrystalline (Cu$_{1-x}$Ag$_x$)$_2$OSeO$_3$ and Cu$_2$(O$_{1-x}$S$_{x}$)$_2$SeO$_2$. \textbf{a-d)} (Cu$_{1-x}$Ag$_x$)$_2$OSeO$_3$ is shown on the left while \textbf{e-h)} Cu$_2$(O$_{1-x}$S$_{x}$)$_2$SeO$_2$ is shown on the right. $x=0.05$ is shown on top while $x=0.10$ is shown on the bottom. The skyrmion region is broadened by increased substitution in both the Ag and S cases. Each SANS rotation plot represents the highest symmetry rotation pattern achieved after rotation around the vertical axis across a span of 180 degrees approximately five times. \add{Figures \textbf{c)}, \textbf{d)}, \textbf{g)}, and \textbf{h)} were integrated for 300 s, 220 s, 330 s, and 140 s, respectively.} Temperature and field location of each rotation is indicated by a red square labeled with roman numerals i-iv corresponding to figures \textbf{c)}, \textbf{d)}, \textbf{g)}, and \textbf{h)}, respectively. Bars on the mark represent range of possible measurement conditions. In all cases, six-fold symmetry is observed, though increased substitution decreases the six-fold intensity. Phase boundaries were determined as they were for Cu$_2$OSeO$_3$ in Fig. \ref{fig:parentchar}.}
	\end{figure*}
	
	We use the following procedure to distinguish single-q helices from skyrmion or multi-q helical magnetic structures. SANS measurements are taken with the magnetic field aligned with the neutron beam. Thus, only helical order perpendicular to the field direction (ie, skyrmions or mult-q helices) is measured. The result is a peak in SANS intensity at $|q| \approx 0.01\ \text{\AA}^{-1}$ with no azimuthal dependence. Importantly, this magnetic peak sits atop a large nuclear background in a polycrystal due to power-law scattering from the polydispersed, micron-scale grains. However, the area under the magnetic peak can still be accurately determined with sufficiently sophisticated fitting (see Appendix B). A plot of the integrated intensity of magnetic scattering due to in-plane helical order as a function of temperature and field is shown in Fig. \ref{fig:parentchar}(b) for parent polycrystalline Cu$_2$OSeO$_3$. A peak in intensity is apparent at $57\ K$ and $25\ mT$, in agreement with the magnetization map of Fig. \ref{fig:parentchar}(a)  (within the uncertainty of the calibration of the magnet used in SANS) and published results \cite{onose2012observation,seki2012formation,seki2012observation}. At low field and low temperature, a weak signal is observed which can be associated with the onset of the multi-q helical phase. This analysis demonstrates that the SANS technique is able to distinguish the phases with in-plane helices---including the skyrmion phase---from the surrounding phases, even in a powdered sample (the same measurement on (Cu$_{0.90}$Zn$_{0.10}$)$_2$OSeO$_3$ is presented in Appendix C).

	We also use a technique developed previously \cite{gilbert2019precipitating} to collectively orient the skyrmion lattices, thus distinguishing a hexagonal skyrmion phase from other helical phases oriented orthogonal to the field. In performing this proceedure, a polycrystalline sample of Cu$_2$OSeO$_3$ is prepared in a temperature/magnetic field environment which facilitates skyrmion formation, then the sample is rotated about the vertical axis in the magnetic field. After the rotation, the ring feature located at $|q| \approx 0.01\ \text{\AA}^{-1}$ develops six-fold azimuthal symmetry, indicating long-range orientation of the skyrmion lattice. This can be seen in Fig. \ref{fig:rotationGuide}, where six-fold azimuthal symmetry is resolved in polycrystalline Cu$_2$OSeO$_3$ only after rotation in field. To highlight the symmetry, the data are fit [black line in Fig. \ref{fig:rotationGuide}(c, d)] with six coupled Gaussian peaks:
	\begin{equation}
	\begin{split}
	I(\phi) &= B + \sum_{n=1}^{3} A_n \\
	& \times \Biggl[ e^{ -\frac{ { \left[ \left(\phi-\phi_0\right)-\frac{n\pi}{3} \right] }^2 } { 2 w_\phi } }  + e^{ -\frac{ { \left[ \left(\phi-\phi_0\right)-\frac{n\pi}{3}-\pi \right] }^2 } { 2 w_\phi } } \Biggr]
	\end{split}
	\label{eq:rotFit}
	\end{equation}
	where $B$ is a constant background, $\phi$ is the azimuthal coordinate, $\phi_0$ is an azimuthal offset, $w_\phi$ is the width of the peak, and the three $A_n$ are the height of each pair of opposing Gaussians. This same work also showed that rotating the sample outside of the skyrmion window resulted in a two-peak or uniform pattern.
	
	\add{Though not the focus of this inves1tigation, time dependent relaxation of the six-fold symmetry after rotation was observed in the substituted samples. The relaxation was observable on the timescale of minutes, consistent with previous observations of similar time-dependent lattice relaxations \cite{bannenberg2017reorientations}. In the rotation measurements of substituted samples, scattering intensity was collected in bins of $\approx 30$ s, and integrated for the time indicated in the caption, starting $\approx 10$ s after rotation. The number of bins to integrate over was chosen to maximize the clarity of the six-fold symmetry, as longer counting improves statistics, although the peak intensity decreases with time.}
	
	The intensity mapping of Fig. \ref{fig:parentchar} allows direct confirmation of helical order perpendicular to magnetic field, while the rotation technique of Fig. \ref{fig:rotationGuide} allows confirmation of six-fold symmetry. By comparing both measurements to magnetization measurements, the magnetic phase diagram can be effectively mapped in polycrystalline samples. In what follows, these techniques are used together to characterize the magnetic structure of various chemical substitutions of Cu$_2$OSeO$_3$.

	\subsection{Silver and Sulfur Substitutions}
	Silver is substituted into the parent compound by substituting AgO for CuO in the initial composition, forming (Cu$_{1-x}$Ag$_x$)$_2$OSeO$_3$ with $x=0.05$ and $x=0.10$. The magnetic phase diagrams as determined by magnetometry ($dM/dH$) are shown in Fig. \ref{fig:SAgMPMS}(a) and (b), while SANS data taken after the sample is rotated are shown in Fig. \ref{fig:SAgMPMS}(c) and (d) for  $x=0.05$ and $x=0.10$, respectively. Both substitutions show qualitatively similar results to the parent compound. The onset of magnetic order remains at $\approx 58$ K. Additionally, both samples show enhancement of the low-field, multi-q helical phase stability with respect to magnetic field and a decrease in the saturation field. Sulfur is then substituted into the parent compound by replacing CuO with CuS to create Cu$_2$(O$_{1-x}$S$_{x}$)$_2$SeO$_2$ with $x=0.05$ and $x=0.10$. The magnetic phase diagrams are again shown in Fig. \ref{fig:SAgMPMS}(e) and (f), while SANS data taken after the sample is rotated are shown in Fig. \ref{fig:SAgMPMS}(g) and (h) for  $x=0.05$ and $x=0.10$, respectively. The onset of magnetic order again remains relatively unchanged at $\approx 58$ K. As in the silver substitutions, the sulfur substitutions show similar enhancement of the low-field, multi-q helical phase, and suppression of the saturation field. Saturation magnetization as a function of temperature is shown in Appendix A. In effect, chemical substitution of Ag or S appears to shrink the single-q helical phase region.
	
	Remarkably, the $dM/dH$ dip associated with skyrmion order becomes extended in temperature as both the Ag and the S substitution level is increased. The high temperature boundary of the skyrmion stability envelope remains relatively unchanged with doping, but the skyrmion region extends to larger fields and lower temperatures with substitution. In both the Ag and S substitutions, the temperature range spans 10 K at $x=0.10$. SANS measurements are consistent with the magnetometry results, showing 
	six-fold symmetry after rotation in the skyrmion window [Fig. \ref{fig:SAgMPMS}(c,d) and (g-h) for the Ag and S substitutions, respectively]. SANS six-fold symmetry is plotted relative to the nuclear scattering contribution (see Appendix B). Note that the $x=0.05$ sulfur substitution shows larger magnetic scattering (grey) than other substitutions: much closer to the parent. This suggests that the sulfur has a weaker effect on the magnetic order compared to other substitutions. Additionally, the skyrmion SANS signature is strongest towards the high temperature boundary of the skyrmion stability envelope, consistent with previous reports \cite{seki2012formation, makino2017thermal}. However, as evidenced in Fig. \ref{fig:SAgMPMS}, increased substitution decreases the skyrmion lattice signature compared to the parent (as confirmed later in Fig. \ref{fig:skyrmion-ness}). Decreased scattering intensity can be the result of a reduced magnetic moment, or a change in the scattering form or structure factors. We expect here the decreased intensity to be the result of a reduced moment and structure factor. The reduced structure factor indicates increased disorder within the hexagonal skyrmion lattice, potentially due to local variations in the skyrmion pitch, pinning of the skyrmions during the lattice formation, or a coexistence of non-lattice skyrmions and helices.
	
	These results show that both Ag and S substitution cause similar changes in the skyrmion phase diagram. We suggest that the underlying mechanism for this broadening is chemical disorder. In the case of the Ag substitution, the behavior could also be caused by change of valence or weakening of magnetic interaction strength as the magnetically active copper site is replaced. However, if the broadening were due to magnetic interactions, the sulfur substitution should not show the same behavior. Another alternative is that the substituted sample possesses distinct phase separation or stoichiometric inhomogeneity. However, measuring the saturation magnetization as a function of temperature shows qualitatively similar behavior to the parent without any additional inflections (see Appendix A), as would be expected for a mixed phase material. This suggests that the sample is single phase and chemically homogeneous. The broadening could also be due to chemical expansion as the substitutions expand the lattice. However, previous work has shown that positive applied pressure expands the skyrmion stability envelope while negative chemical pressure contracts it \cite{wu2015physical} in contradiction to the behavior seen here. Therefore, chemical disorder is the most likely cause.

	\subsection{Zinc Substitution}
	
\begin{figure}
		\includegraphics[width=\columnwidth]{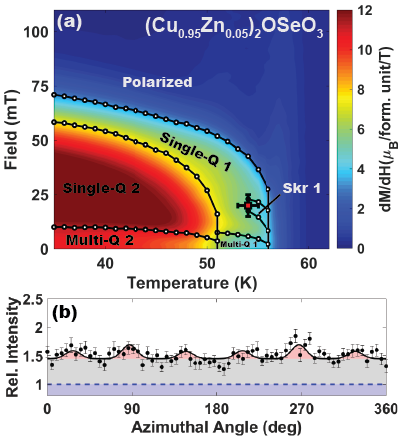}
		
		\caption{\label{fig:Zn5pMPMS} Skyrmion phase in (Cu$_{0.95}$Zn$_{0.05}$)$_2$OSeO$_3$. \textbf{(a)} shows the $dM/dH$ magnetization contour plot for (Cu$_{0.95}$Zn$_{0.05}$)$_2$OSeO$_3$. The phase boundaries, determined from the magnetization in the same way as the previous samples, are indicated with black lines.
		\textbf{(b)} shows an annular cut of a SANS measurement of the sample at the point indicated with a red square in \textbf{(a)} (bars on the mark represent range of possible measurement conditions). \textbf{(b)} is performed at 54 K and 20 mT \add{for 240 s immediately after rotation}, and shows weakened six-fold symmetry compared to the parent compound (see Fig. \ref{fig:rotationGuide}).}
	\end{figure}

\begin{figure}
		\includegraphics[width=\columnwidth]{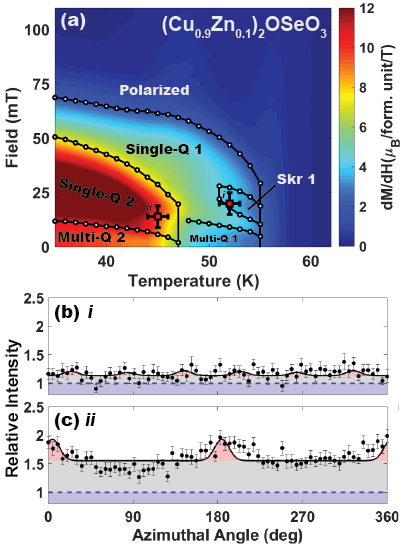}
		
		\caption{\label{fig:Zn10pMPMS} Skyrmion phase in (Cu$_{0.90}$Zn$_{0.10}$)$_2$OSeO$_3$. \textbf{(a)} shows the $dM/dH$ magnetization contour plot for (Cu$_{0.90}$Zn$_{0.10}$)$_2$OSeO$_3$. The phase boundaries, determined from the magnetization in the same way as the previous samples, are indicated with black lines.
		\textbf{(b)} and \textbf{(c)} show annular cuts of SANS measurements of the sample at the conditions---labeled i and ii, respectively---indicated with red squares in \textbf{(a)} (bars on the mark represent range of possible measurement conditions). \textbf{(b)} is performed at 52 K and 20 mT \add{for 300 s after rotation} while \textbf{(c)} is perfomed at 45 K and 14 mT \add{for 420 s after rotation}. \textbf{(b)} shows weakened six-fold symmetry compared to the parent compound or (Cu$_{0.95}$Zn$_{0.05}$)$_2$OSeO$_3$ (see Fig. \ref{fig:rotationGuide} and \ref{fig:Zn5pMPMS}). \textbf{(c)} shows \add{no apparent}\rmv{ the lack of an} ordered skyrmion lattice\add{, instead favoring}\rmv{ in favor of} helical, two-fold order.}
	\end{figure}

\begin{figure}
		\includegraphics[width=\columnwidth]{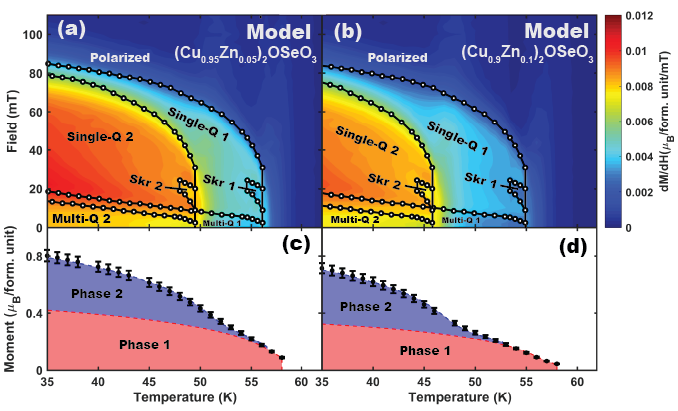}
		
		\caption{\label{fig:ZnModel} Modeling the magnetization of (Cu$_{1-x}$Zn$_x$)$_2$OSeO$_3$. \textbf{(a)} and \textbf{(b)} attempt to replicate the actual magnetization of (Cu$_{1-x}$Zn$_x$)$_2$OSeO$_3$ with the superposition of two identical, weighted, temperature-shifted, magnetization measurements for Cu$_2$OSeO$_3$ as described in Equation \ref{eq:Znfit} with parameters given in Table \ref{tab:ZnParams}. Phase boundaries for each data-set are indicated with black boundaries. This modeling of the Zn magnetization replicates many of the important features found in Fig. \ref{fig:Zn5pMPMS}(a) and \ref{fig:Zn10pMPMS}(a), and suggests that the Zn substitutions are well described by two partial volume fractions of different stoichiometries, creating two different superimposed magnetic phase diagrams. \textbf{(c)} and \textbf{(d)} show the real magnetization of each Zn sample (black) as a function of temperature at a constant field of 100 mT. There are two magnetic transition temperatures as the sample transitions from paramagnetic to field-polarized. The contribution to $M$ from the first model phase is shown in red, while the contribution from the second model phase is shown in blue. The sum of the two model phases (red and blue) closely resembles the actual magnetization (black).}
	\end{figure}

	Polycrystalline (Cu$_{1-x}$Zn$_x$)$_2$OSeO$_3$, as previously reported \cite{wu2015unexpected,vstefanvcivc2018origin}, shows strikingly different behavior from (Cu$_{1-x}$Ag$_x$)$_2$OSeO$_3$ or Cu$_2$(O$_{1-x}$S$_{x}$)$_2$SeO$_2$. Zn substitution creates a coexistence of two distinct stoichiometr\add{ies}\rmv{ ic phases} within these powdered samples \add{(as supported by previous reports of the different scoichiometries producing multiple nearby peaks in high-resolution X-ray powder diffraction \cite{vstefanvcivc2018origin})}. Both stoichiometr\add{ies}\rmv{ ic phases} demonstrate similar heli-magnetic phase diagrams, but the nominally higher Zn concentration phase has a signifigantly lower transition temperature than the lower Zn concentration phase. The $dM/dH$ phase diagram of $x=0.05$ is shown in Fig. \ref{fig:Zn5pMPMS}(a), and of $x=0.10$ is shown in Fig. \ref{fig:Zn10pMPMS}(a). Note that the $x=0.05$ sample shows transitions at approximately $51\ K$ and $56\ K$ while the $x=0.10$ sample shows transitions at approximately $47\ K$ and $55\ K$. Thus, increasing Zn concentration lowers the transition temperature in both stoichiometr\add{ies}\rmv{ ic phases}. The coexistence of these two magnetic transitions is clearly manifested as a kink in the magnetization as a function of temperature [Fig. \ref{fig:ZnModel}(c-d)].
	To further establish that the magnetization of the Zn substituted samples is well explained by the coexistence of two stoichiometr\add{ies}\rmv{ ic phases}, we use the magnetization data of Cu$_2$OSeO$_3$, $M_{parent}(T,H)$\add{, which exhibits the largest saturation magnetization (see black curve in Appendix A, Fig. \ref{fig:sat_mag}(a))}, as a functional form to qualitatively model the Zn substituted magnetization data, $M_{Zn}(T,H)$ \add{(also shown in Fig. \ref{fig:sat_mag}(a))}. The shifting transition temperatures and the volume fraction of each stoichiometr\add{y}\rmv{ ic} phase within a sample are accommodated in the model as
	\begin{equation}
	M_{Zn}(T,H) \approx c_1 M_{parent}(T-T_1,H) + c_2 M_{parent}(T-T_2,H).
	\label{eq:Znfit}
	\end{equation}
	$T_i$ is the transition temperature shift and $c_i$ is the fractional contribution, relative to the parent saturation magnetization, of each stoichiometr\add{y}\rmv{ ic phase} within the sample. $M_{parent}(T,H)$ is used as a model to emphasize the distinct phases, not to capture all details of the phase diagram perfectly. The model is fit to the data along $M_{Zn}(T,H=100\ mT)$ \add{(shown in Fig. \ref{fig:sat_mag}(a))} so as to consider only\rmv{ the field-polarized transitions} \add{the transition from the paramagnetic to the field-polarized state in each of the two distinct stoichiometries within the sample} and not the more nuanced helimagnetic phases\rmv{ ($M(T,H=100\ mT)$ for each sample is shown in Fig. \ref{fig:sat_mag}(a))}. These fits of $M_{Zn}(T,H=100\ mT)$, shown in Fig. \ref{fig:ZnModel}(c-d), obtain reduced $\chi^2$ of 0.22 and 0.45 for $x_{Zn}=0.05$ and $x_{Zn}=0.10$, respectively. The fit parameters are given in Table \ref{tab:ZnParams}.
	\begin{table*}[!hbt]
		\caption{\label{tab:ZnParams} Parameters used to fit (Cu$_{1-x}$Zn$_x$)$_2$OSeO$_3$ magnetization data to model Cu$_2$OSeO$_3$ magnetization data as per Eq. \ref{eq:Znfit}. Uncertainties here represent a 95\% confidence interval. Results of fit for each value of $x_{Zn}$ shown in Fig. \ref{fig:ZnModel}.}
		\begin{ruledtabular}
			\begin{tabular}{ldddd}
				\textrm{$x_{Zn}$}&
				\textrm{$c_1$}&
				\textrm{$T_1$}&
				\textrm{$c_2$}&
				\textrm{$T_2$}\\
				\colrule
				0.05 & 0.48 \pm 0.03 & -1.9 \pm 0.2 & 0.46 \pm 0.04 & -8.5 \pm 0.4\\
				0.10 & 0.37 \pm 0.02 & -3.1 \pm 0.2 & 0.52 \pm 0.03 & -12.2 \pm 0.5\\
			\end{tabular}
		\end{ruledtabular}
	\end{table*}
	These parameters characterize the volume fraction and temperature shifts of two distinct phases within (Cu$_{1-x}$Zn$_x$)$_2$OSeO$_3$.

	The magnetic phase diagrams in $dM/dH$ generated from this model of $M(T,H=100\ mT)$ are shown in Fig. \ref{fig:ZnModel}(a-b). The agreement between these phase diagram models and the data in Fig. \ref{fig:Zn5pMPMS}(a) and Fig. \ref{fig:Zn10pMPMS}(a) is excellent, with the exception of the low temperature skyrmion region as discussed later.  Thus these polycrystalline, Zn-substituted samples can be treated as containing two distinct stoichiometr\add{ies}\rmv{ ic phases} with different Zn concentrations, leading to a superposition of two helimagnetic phase diagrams with separate transition temperatures dictated by the Zn concentration (with the lower temperature phase likely containing more Zn). These two stoichiometries must have the same crystal structure but with different Zn substitution levels, as X-ray powder diffraction does not reveal a second crystal structure within the sample. It is possible that the segregation of regions with different Zn substitution levels is as small as sub-micron, however spatially resolved compositional analysis would be required to determine the real-space distribution of these domains. Admittedly, this model does not fully capture all details of the magnetic phase diagram below the field polarized state---particularly the lower saturation field of the substituted samples. However, using Cu$_2$(O$_{0.95}$S$_{0.05}$)$_2$SeO$_2$ as a model in the same way as above (not shown) results in even better modelling of the magnetic phase diagram, suggesting there are similar chemical disorder-induced effects in the Zn phase diagram, as was noted for the Ag and S substitutions.

\begin{figure}
		\includegraphics[width=\columnwidth]{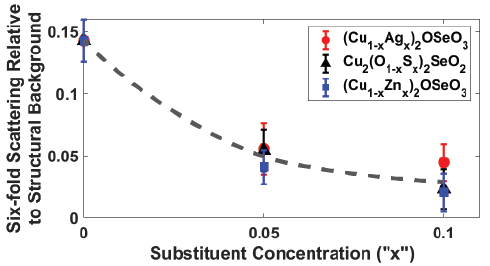}
		
		\caption{\label{fig:skyrmion-ness} Scaled skyrmion lattice SANS scattering intensity. The ratio between the six-fold symmetric scattering intensity and the structural scattering intensity is plotted as a function of concentration for (Cu$_{1-x}$Zn$_x$)$_2$OSeO$_3$, (Cu$_{1-x}$Ag$_x$)$_2$OSeO$_3$, and Cu$_2$(O$_{1-x}$S$_{x}$)$_2$SeO$_2$. The six-fold symmetric scattering can be attributed to the skyrmion phase. The dashed line is a guide to the eye. There is a general, monotonic decrease in skyrmion scattering intensity as substituent concentration is increased. Details of the calculation are given in Appendix B.}
	\end{figure}

	The rotation method (see Fig. \ref{fig:rotationGuide} and associated discussion) allows further characterization of these skyrmion phases. Figure \ref{fig:Zn5pMPMS}(b) and \ref{fig:Zn10pMPMS}(b, c) show annular cuts of SANS data after rotation in the higher temperature candidate skyrmion phase in (Cu$_{0.95}$Zn$_{0.05}$)$_2$OSeO$_3$ and the high and low temperature candidate skyrmion phases in (Cu$_{0.90}$Zn$_{0.10}$)$_2$OSeO$_3$, respectively. Hexagonal symmetry is shown in both higher temperature skyrmion phases, but weakens as Zn concentration is increased. This can be associated with the increased substitution, as noted in Ag and S, and also with the volume fraction of the sample that contributes to the phase, which decreases significantly from Fig. \ref{fig:Zn5pMPMS}(b) to \ref{fig:Zn10pMPMS}(b) (see Table \ref{tab:ZnParams}). Indeed, plotting the integrated intensity of hexagonal scattering relative to the nuclear scattering as a function of substitution shows an overall decrease in the relative magnetic scattering intensity (Fig. \ref{fig:skyrmion-ness}).
	In contrast, rotating the (Cu$_{0.90}$Zn$_{0.10}$)$_2$OSeO$_3$ sample in the lower temperature candidate skyrmion phase reveals a completely different behavior. After the rotation there is no \add{apparent} six-fold symmetry, but rather two-fold symmetry across the rotation axis indicating helical order perpendicular to the rotation axis. This suggests that \add{rotation in} the proposed lower temperature skyrmion \add{region promotes a helical state}\rmv{ lattice phase, if it exists at all, is highly unstable and can be transitioned to a predominately helical state by rotation (n}\add{. N}ote that it is possible that there is still a coexistence of non-lattice skyrmions \add{or a dominance of the magnetocrystalline anisotropy in determining the skyrmion lattice orientation}\rmv{)}.

	The fact that SANS shows helical order perpendicular to the magnetic field upon rotation in the lower temperature candidate skyrmion region also suggests that the helices are generally not as strongly aligned with the field direction as the helices in the parent Cu$_2$OSeO$_3$, where rotation aligns all helices with the field.

	\section{Conclusion}
	
	Here, we present results using magnetometry and SANS to study the effects of substitution at $x=\left \{0.05,0.10\right \}$ in polycrystalline (Cu$_{1-x}$Zn$_x$)$_2$OSeO$_3$, (Cu$_{1-x}$Ag$_x$)$_2$OSeO$_3$, and Cu$_2$(O$_{1-x}$S$_{x}$)$_2$SeO$_2$. We demonstrate the use of SANS to resolve skyrmion order in polycrystalline samples. While magnetization measurements indicate that Ag and S substitution enhances the temperature stability of the skyrmion phase, SANS measurements show reduced ordering of the skyrmion lattice with substitution. Additionally, while we show that the higher temperature stiochiometry in (Cu$_{0.90}$Zn$_{0.10}$)$_2$OSeO$_3$ contains a skyrmion lattice phase, \add{we do not see evidence of} the second, lower temperature candidate skyrmion region\rmv{ favors the helical phase}. Chemical disorder induces disorder in the skyrmion lattices of Cu$_2$OSeO$_3$. This could manifest as a coexistence of skyrmion and helical order as has been seen in thin films (Ref. \cite{yu2010real}, for example), or as a destruction of helical order entirely in parts of the system. Magnetic microscopy would prove enlightening to understanding the actual microstructure of these substitutions. Additionally, synthesis of single crystals or thin films with significant chemical substitution are still a worthy endeavor, and could improve sample homogeneity and reduce lattice disorder. Overall, these substitutions offer insight into the stability of skyrmion order in Cu$_2$OSeO$_3$, and may aid future developments in the technological applications of magnetic skyrmions.

	\begin{acknowledgments}
		The authors would like to thank Paul Kienzle for support regarding SANS fitting; Jeffrey Krzywon, Tanya Dax, and Qiang (Alan) Ye for assistance with SANS instrumentation; Kefeng Wang and Halyna Hodovanets for help with growth procedures; and Juscelino Leão for useful discussion of data analysis. Support for Paul Neves was provided by the Center for High Resolution Neutron Scattering, a partnership between the National Institute of Standards and Technology and the National Science Foundation under Agreement No. DMR-1508249. Additionally, we acknowledge the support of the National Institute of Standards and Technology, U.S. Department of Commerce, in providing the neutron research facilities used in this work. Certain commercial materials are identified in this paper to foster understanding. Such identification does not imply recommendation or endorsement by the National Institute of Standards and Technology, nor does it imply that the materials or equipment identified are necessarily the best available for the purpose. The authors report no conflicts of interest in this investigation.
		
	\end{acknowledgments}
	

	\appendix
	\section{Saturation Magnetization}
\begin{figure}[t]
		\includegraphics[width=\columnwidth]{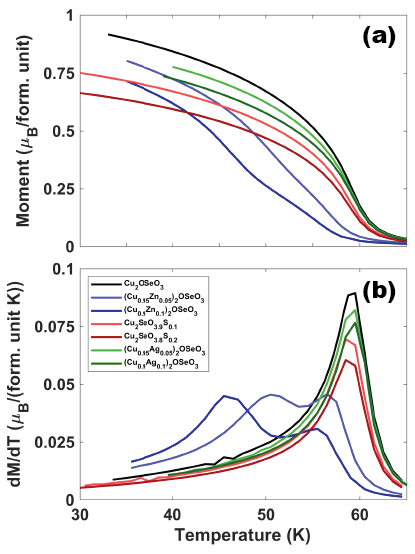}
		
		\caption{\label{fig:sat_mag} Temperature, $T$, dependence of magnetization, $M$ in the field-polarized state. (a) $M$ of each substitution as a function of $T$ at a constant field of 100 mT. (b) The derivative of (a) with respect to temperature, $dM/dT$.}
	\end{figure}

	Magnetization, $M$, as a function of temperature, $T$, is shown for each substitution studied above in Fig. \ref{fig:sat_mag}(a). All magnetization measurements were taken at 100 mT (during field sweeps) so as to be in the field-polarized state. The parent compound shows the highest saturation magnetization, while increasing substitution lowers the saturation magnetization. This lowering saturation magnetization is expected when substituting the magnetically active site with nominally non-magnetic Ag and Zn. The derivative of $M$ with respect to $T$, $dM/dT$, is given in Fig. \ref{fig:sat_mag}(b). A peak in $dM/dT$ corresponds to the magnetic transition temperature. The Ag and S substitutions show little change in the onset of magnetic order. However, the zinc substitutions show two peaks in $dM/dT$ for the two stoichiometr\add{ies}\rmv{ ic phases} as discussed above and in previous work \cite{vstefanvcivc2018origin}.

	\section{SANS Analysis}
	Here, we outline the analysis that leads to our results on skyrmion SANS intensity as presented in Fig. \ref{fig:parentchar}(b), \ref{fig:skyrmion-ness}, \ref{fig:znSANSmap}, and the related discussion.
	
\begin{figure}[t]
		\includegraphics[width=\columnwidth]{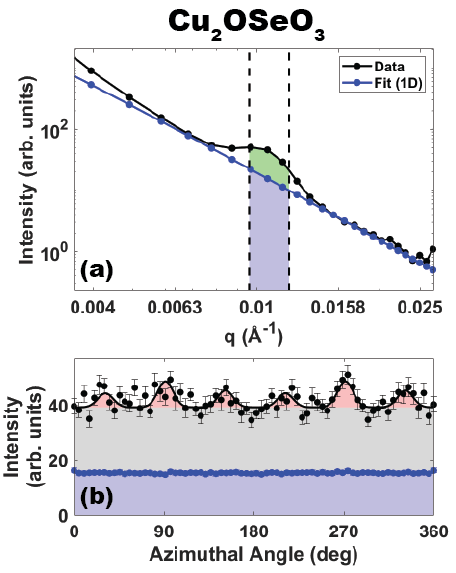}
		
		\caption{\label{fig:SANSfitguide} Guide to SANS analysis. Data for polycrystalline parent Cu$_2$OSeO$_3$ in the skyrmion state at 57 K after rotation [as shown in Fig. \ref{fig:rotationGuide}(b)] used here for clarity. \textbf{(a)} shows intensity of SANS as a function of $q$ (black). A large structural background due to the powder granules is fit with a model function (blue). This allows isolation of the magnetic peak integrated intensity centered near $|q| \approx 0.01\ \text{\AA}^{-1}$. \textbf{(b)} shows SANS intensity as a function of azimuthal angle within the $q$ range indicated by black dashed vertical lines in \textbf{(a)}. Rotation of the sample in magnetic field precipitates six-fold symmetry in the data (black), while the model structural scattering leads to a uniform, lower intensity background (blue). The structural scattering intensity is shaded light blue, uniform magnetic scattering is shaded grey, and six-fold symmetric magnetic scattering is shaded red. The ratio of six-fold symmetric integrated intensity to structural background integrated intensity is a figure of merit for skyrmion lattice order.}
	\end{figure}

	To generate the field-perpendicular helimagnetic scattering intensity maps shown in Fig. \ref{fig:parentchar}(b) and \ref{fig:znSANSmap}, SANS measurements were performed at a variety of temperature and field conditions. The sample was saturated in\rmv{ field} \add{a field of $\approx 0.3$ T} to the field polarized state after each temperature change\rmv{ to remove hysteretic effects}. The black $I(q)$ plotted in Fig. \ref{fig:SANSfitguide}(a) shows exemplary helimagnetic scattering observed in the parent skyrmion phase. Each SANS $I(q)$ dataset is fit with a Markov Chain Monte Carlo fitting algorithm using Bumps \cite{kienzle2011bumps,vrugt2009accelerating}. A Guinier-Porod function \cite{hammouda2010new} is used to empirically model the powder grain nuclear scattering, and a Gaussian peak is added to empirically model the helimagnetic scattering centered at $|q| \approx 0.01\ \text{\AA}^{-1}$. Then, the integrated intensity of the Gaussian peak is plotted for each $I(q)$ as a function of temperature and field. This maps the field-perpendicular helimagnetic scattering as a function of temperature and field.
	
	To generate Fig. \ref{fig:skyrmion-ness}, we performed a separate but related analysis of the SANS pattern after rotation. First, we modeled the nuclear scattering in $I(q)$ with a power-law function [blue in Fig. \ref{fig:SANSfitguide}(a)]. (The Guinier-Porod nuclear scattering model is unnecessary here as the magnetic scattering at the center of the skyrmion phase after rotation is sufficiently strong to allow for a less sophisticated model.) The data (black) and model (blue) integrated intensity within the selected skyrmion $q$ range [indicated in Fig. \ref{fig:SANSfitguide}(a) with vertical dashes] is plotted as a function of azimuthal angle in Fig. \ref{fig:SANSfitguide}(b). These $q$ bounds are chosen as they generate the clearest six-fold symmetry in the data. The power law model in Fig. \ref{fig:SANSfitguide}(a) is used to determine the azimuthally uniform, nuclear contribution (blue points) to the signal in Fig. \ref{fig:SANSfitguide}(b), and allows calculation of the nuclear scattering integrated intensity within this $q$ range. The data (black points) show six-fold symmetric scattering from skyrmion order precipitated by the rotation (red), and isotropic helimagnetic and non-lattice skyrmion scattering (grey) on top of the nuclear scattering (blue). The data are fit with Equation \ref{eq:rotFit} to obtain the integrated intensity of the six-fold symmetric scattering and the non-six-fold scattering. Then, the ratio of the six-fold scattering intensity to the nuclear scattering intensity [red area to blue area in Fig. \ref{fig:SANSfitguide}(b)] can be plotted as a function of substituent concentration (Fig. \ref{fig:skyrmion-ness}). This ratio offers a figure of merit for the amount of scattering due to ordered skyrmion lattices, normalized by the nuclear background. This normalizes intensity scaling due to varied sample mass in the neutron beam among samples.
	
	\add{It is also possible to isolate the magnetic scattering from the nuclear scattering by subtracting a high temperature or high field background measurement, then fitting the area of the peak centered at $|q| \approx 0.01\ \text{\AA}^{-1}$. Both directly fitting the nuclear scattering and subtracting a background measurement give qualitatively similar results. However, this background subtraction method does not isolate the helimagnetic scattering quite as well as the fitting method presented here. There is a field dependent contribution to the background scattering, as spins in each granule paramagnetically align with the field, that adds to the nuclear powder scattering in directions perpendicular to the magnetic field. Simple background subtraction does not fully capture all the details of this magnetic field-dependent scattering.}

	\section{Zn Substitution SANS Magnetic Scattering}
\begin{figure}
		\includegraphics[width=\columnwidth]{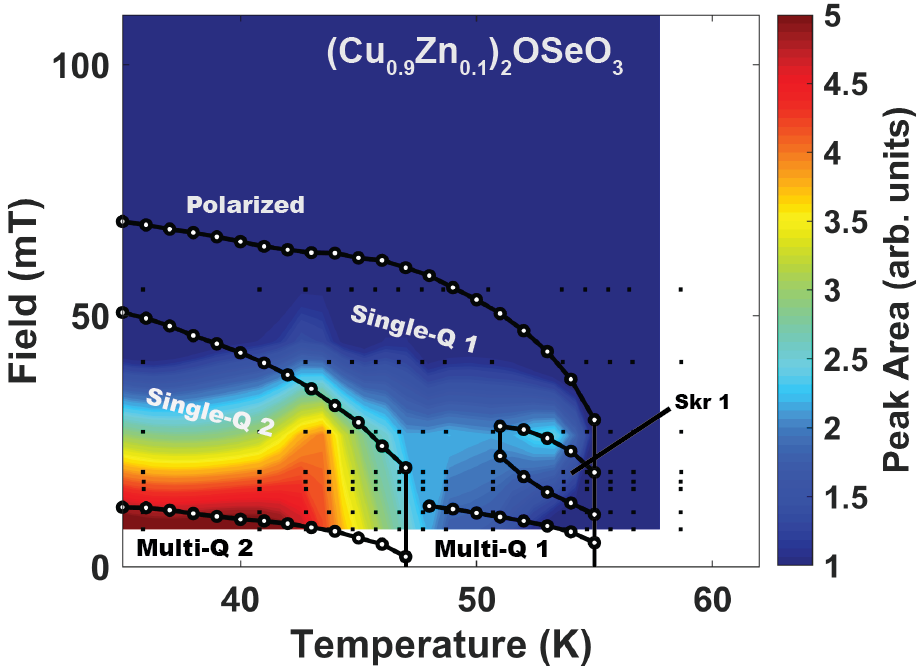}
		
		\caption{\label{fig:znSANSmap} $|q| \approx 0.01\ \text{\AA}^{-1}$ magnetic peak SANS integrated intensity map of helical order perpendicular to applied magnetic field as a function of temperature and field strength in polycrystalline (Cu$_{0.90}$Zn$_{0.10}$)$_2$OSeO$_3$ [analogous to Fig. \ref{fig:parentchar}(b)]. Field is applied parallel to the incident neutron beam. Measurements are taken at many temperatures and fields (shown with black points). Integrated scattering intensity due to the $|q| \approx 0.01\ \text{\AA}^{-1}$ magnetic peak corresponding to out-of-plane helical order is determined from fitting the magnetic peak on top of a structural, power-law background. Phase boundaries determined from magnetization measurements [Fig. \ref{fig:Zn10pMPMS}(a)] are shown. A maximum in intensity is seen corresponding to Skr 1 at 53 K and 27 mT. Large, field-perpendicular helical intensity is also visible at low temperature and field due to the helical phase. No peak in intensity can be resolved corresponding to the \add{reported lower temperature skyrmion}\rmv{ Skr 2} region.}
	\end{figure}

	The magnetic phase diagram for polycrystalline (Cu$_{0.90}$Zn$_{0.10}$)$_2$OSeO$_3$ in Fig. \ref{fig:Zn10pMPMS}(a) is supported by a mapping (Fig. \ref{fig:znSANSmap}) of the SANS field-perpendicular helimagnetic scattering at $|q| \approx 0.01\ \text{\AA}^{-1}$, as was done for the parent Cu$_2$OSeO$_3$ in Fig. \ref{fig:parentchar}(b). Fig. \ref{fig:znSANSmap} shows a peak in field-perpendicular helimagnetic scattering at 53 K and 25 mT corresponding to Skr 1.
	This plot also shows strong field-perpendicular helical order in the low temperature single-q helical phase in comparison to the parent single-q helical phase. This \add{is consistent with}\rmv{ agrees with} the behavior of (Cu$_{0.90}$Zn$_{0.10}$)$_2$OSeO$_3$ SANS under field rotations in this region [Fig. \ref{fig:Zn10pMPMS}(c)]. This suggests that disorder causes the helices in the second, lower temperature helical phase to be less polarized along the field direction, leading to greater field-perpendicular helimagnetic ordering. \add{It is possible that the increased intensity at lower temperatures could be due partially to a second skyrmion stability window as has been recently reported \cite{chacon2018observation, bannenberg2019multiple, qian2018new}, however, no hexagonal pattern was observed in the SANS data before or after rotation in this region.}\rmv{ If a second, lower temperature skyrmion phase exists at all, it coexists with dominant multi-q helical structures.} \add{Additionally,}\rmv{ Note that} the field-perpendicular helical order below the second transition is much stronger than that in the Skr 1 region because Skr 1 is only present in \add{of order one}\rmv{ approximately} half \add{of} the sample (see Table \ref{tab:ZnParams}), while the helical order is \add{nominally} full volume fraction below the second transition. The SANS intensity due to the skyrmions is thus about half as intense as the full-volume helical intensity, as expected.

	%

\end{document}